# Observation of sharp metamagnetic transition, Griffiths like phase and glassy nature in double perovskite $Eu_2CoMnO_6$

**Mohd Alam[1], Arkadeb Pal[1], Khyati Anand[1], Prajyoti Singh[1] and Sandip Chatterjee[1*]**

[1]*Indian Institute of Technology (B.H.U.), Varanasi – 221005, (U.P.) India.*

***Corresponding author: schatterji.app@itbhu.ac.in**

## Abstract

In the present investigation, some novel magnetic behaviors exhibited by double perovskite (DP) $Eu_2CoMnO_6$(ECMO) has been reported. XRD analysis of ECMO showed that it has a monoclinic crystal structure (space group P $2_1$/n). A second-order magnetic phase transition as a sudden jump in the magnetization curve has been observed at 124.5 K. This is related to the paramagnetic to ferromagnetic/E*-type antiferromagnetic phase transition due to the competing Co-O-Mn exchange interactions. A clear low-temperature compensation point followed by negative magnetization is observed in the zero-field-cooled curve of the sample, suggesting the formation of canted ferromagnetic domains or antiparallel spins and clusters that are separated by an antiphase boundary. The large bifurcation between the ZFC and FC curves has been observed, suggesting strong spin frustration is present in the system. More interestingly, sharp multiple steps in magnetization are observed in M-H curve at 2 K and observed only in the forward field-sweep direction which vanishes on increasing temperature. Moreover, prominent smaller peaks immediately above the long-range ordering temperature are observed suggesting the presence of preformed percolating clusters which eventually gives rise to Griffiths like phase which is seen in DC as well in AC susceptibility. The real part of AC susceptibility with DC bias shows an unusual sharp peak near $T_C$ that broadens on increasing field strength and splits into two maxima around 750 Oe, which is attributed to the presence of critical fluctuations associated with a continuous transition to the FM state and large magnetic anisotropy in the system.

## INTRODUCTION

Having taken into consideration the J. C. Maxwell's equations, interconnecting the dynamics of electric fields, magnetic fields, and electric charges, it was inferred that the magnetic interactions and motion of electric charges are not independent but intrinsically coupled to each other showing unification of magnetism and electricity[1]. So we can say that the magnetic properties can be biased with electrical properties and vice-versa, and the materials having such extraordinary, multi-functional properties are gained much more interest due to their attractive, well-to-do physics and panorama for technological device applications[2-4]. Moreover, one of the most interesting magnetic phenomena in the materials are paying a lot of attention is the field-induced, step-like shape in the isothermal magnetization curves known as a metamagnetic transition (MMT). Metamagnetic transition is a phenomenon in which the system undergoes a transition from a low magnetic-moment to a high magnetic-moment state, mainly belong to a first-order phase transition. The most interesting systems are those in which the presence of strong magneto-structural coupling gives rise to such transition and many other interesting magnetic properties. MMT is observed in certain phase-separated manganites and a few intermetallics like $Gd_5Ge_4$[5-8]. Materials with the coexistence of different magnetic function, such as MMT along with giant magnetocaloric effect have great importance in solid-state cooling technology like high efficient magnetically cooled refrigerator[9-12].

In the last few decades, people are trying to search for new multi-functional materials for next-generation spintronic devices[4, 13]. The rare earth double perovskite (DP) oxides, $A_2BB'O_6$ (A= rare earth ions or alkaline ions; B/B' = transition metal ions) having rock salt ordered structure are of great importance due to their remarkable capacity for multiferroicity. DP's are ferromagnetic insulators driven by the electron correlations possessing high magnetic transition temperatures, close to room temperature. In addition, they display room-temperature magneto-capacitance, giant magneto-resistance, colossal magneto-dielectric, ferroelectricity, spin-glass transition, exchange bias, spin reorientation, local atomic disorder and formation of anti-phase domains, Griffiths phase, the existence of soft phonon coupling with spins, etc[14-21]. Beside experimentally, people have predicted the existence of multiferroicity in DP's like $La_2NiMnO_6$, $Y_2CoMnO_6$, theoretically using first-principles density functional calculations [22-26]. Furthermore, these properties can be tuned by changing valencies of the transition metal, size of cation and environmental or chemical pressure. Moreover, magnetic properties are strongly influenced by the presence of antisite disorder in the system due to the change of B-O-B' exchange interaction path[27, 28]. Ferromagnetism in the DP's is arising mainly due to exchange interaction operating through the paths $B^{2+}$-O-$B'^{4+}$ in a perfect crystallographically ordered system. Although, it is not possible to achieve perfect order in real samples. The samples prepared in a laboratory have antisite disorder due to the interchange of transition metal cation positions. This affects perfect ferromagnetism by introducing $B^{2+}$-O-$B^{2+}$ and $B'^{4+}$-O-$B'^{4+}$ exchange interaction, which causes antiferromagnetic clustered regions[15, 29, 30]. Besides antiferromagnetic clustered regions, compounds intrinsically form antiphase domain (APD) structures due to the different spatial distribution of the B/B′ ordering and B/B or B′/B′ ordering[31].

DP’s having “A” site nonmagnetic ions such as La, Y, Lu, Eu has recently drawn extensive attention due to their distinctive magnetic behavior. As magnetic as well as other physical properties of DP’s are mainly determined by the valence state of the transition-metal cations B and B′ by means of their exchanges interaction with each other and with the rare earth metal ions. When “A” is a nonmagnetic ion then we can neglect interactions between rare-earth cations and transition metal cations. The most intriguing feature observed in $Y_2CoMnO_6$[20,32,33], $Lu_2CoMnO_6$[33] and $Eu_2CoMnO_6$[34] is the step-like shape of the isothermal magnetization curve i.e. MMT. The step positions on the curve are a function of temperature and sweep rate of the external magnetic field and are clearer at a lower temperature. Along with MMT, in the present report we have shown the existence of Griffiths like phase, strong spin frustration and hence gives rise to spin-glass-like behavior, strong magnetic anisotropy giving rise to Hopkinson like peak. In a work, critical behavior near the Curie temperature $T_C$ in the ECMO is systematically investigated. It is reported that critical parameters (β, δ, and γ) in ECMO obtained from K-F plot were found to be 0.6065, 0.9237 and 2.52 with $T_C$ ~130.3 K[35]. These values are very close to critical parameters, which are obtained by the mean-field theory.

## EXPERIMENTAL

Polycrystalline DP ECMO was synthesized by the conventional solid-state reaction method. Well, a mixed stoichiometric mixture of high-purity $Eu_2O_3$, CoO and $Mn_2O_3$ were ground for 40 min and calcined at 1000° C for 24h; after some intermediate grinding, the powder was calcined at 1100° C and 1250° C. The sample was then reground and the powder was hydrostatically pressed into small pellets. The pellets were sintered at 1300° C for 48h and finally slowly cooled (at the 30° C/h) to room temperature. Powder X-ray diffractogram (XRD), Rigaku Miniflex II X-ray diffractometer (Cu-Kα) was used to test out the phase clarity of the samples and was refined by Lebail method using FULLPROF suite software. The magnetic measurements were performed in different applied DC and AC field (at a different frequency) at different temperature by a quantum design magnetic property measurement system (Quantum design-MPMS) superconducting quantum interference device (SQUID) magnetometer. The XPS (X-ray photoemission spectroscopy) study were carried out using an Omicron Multiprobe Surface Science System, equipped with a monochromatic source (XM 1000) and a hemispherical electron energy analyzer (EA 125). While performing the XPS measurement the average base pressure were kept about $3.1 \times 10^{-11}$ Torr with a power of 300 W., estimated. From the analysis of width of the Fermi edge, the total energy resolution was found to be about 0.25 eV for the monochromatic Al Kα line with a photon energy of 1486.70 eV. The pass energy for the wide-surface scan spectrum and for the core level spectrums was maintaned at 50 eV and 30 eV respectively.

## RESULTS AND DISCUSSIONS

The X-ray diffraction of ECMO taken at room temperature with Lebail refinement is shown in fig. 1. From the analysis, it was observed that the crystal Lebail fitting match well to a monoclinic crystal structure with a space group P21/n and no other phases were observed.

Thermal (T) variation of magnetization (M) in zero fields and field cooling (ZFC and FC) of ECMO sample at the different applied field (100 Oe, 1000 Oe, 10000 Oe, 20000 Oe, 30000 Oe) was taken while warming (ZFCW and FCW data) as shown in fig. 2(a). The large bifurcation between the ZFC and FC curves has been observed, and the magnitude of bifurcation increases as applied field increased from 100 Oe to 2T and then it gets decreased at 3T. The differences between M-T curves under the same FC–ZFC applied magnetic fields might be expected due to the presence of spin frustrations which signify a typical spin-glass. However, it is not a sufficient feature of typical spin-glasses, so it alone does not prove spin-glasses behavior, which can also arise from the domain structure of a ferromagnetic material or due to magnetic anisotropy. In order to clarify this issue, we have performed ac susceptibility measurements with different frequencies which are shown in fig. 4, and are discussed later in this paper.

Further, from fig. 2(a) it is clear that ZFC magnetization of the ECMO has only a small narrow maximum with a cusp at a temperature $T_a$ ~123.2 K at 100 Oe and at ~120.3 K at 1000 Oe fields. Further, these maximal's gets broaden at the higher magnetic field (1T, 2T, 3T) and exhibits a plateau-like anomaly along with a shift in cusp temperature ($T_a$) towards lower temperature as magnetic field increases (H), and $T_a$ α H. This might be due to a competition between the local anisotropy and the effect of the external field stand-in on the spins, and the cusp at temperature $T_a$ is the temperature where the two energies are almost of equal magnitude i.e. anisotropy energy and the energy caused by the external field. The bell-like behavior around $T_a$ correlates with the coexistence of low-temperature E*-type AFM & FM and high-temperature E*-type AFM/FM to PM transformations in the sample. Through Curie-Weiss (CW) analysis of FC magnetization using the CW relation,

$$\frac{1}{\chi} = \frac{T}{C} - \frac{T_{cw}}{C} \qquad \text{---(1)}$$

and

$$C = {N\mu_{eff}^2}/{3K_B} \qquad \text{---(2)}$$

We found an effective paramagnetic moment value of $\mu_{eff} = 7.7\mu_B$, and Curie-Weiss temperature, $T_{CW}$ was found to be around 98 K, implying a dominant ferromagnetic phase due to $Co^{2+}$-$O^{2-}$-$Mn^{4+}$ exchange interactions in the sample while few anti-ferromagnetic phase due to $Co^{2+}$-$O^{2-}$- $Co^{2+}$ and $Mn^{4+}$-$O^{2-}$-$Mn^{4+}$ which can be explain by the Goodenough-Kanamori rules[27-28]. Further $T_{CW} < T_C$ which is insignificant if we consider it due to some quantum fluctuation, so it might be due to competition between FM and AFM exchange interactions. The magnetization data for T> 200K (~$2T_{CW}$) was used for the CW analysis.

A clear low-temperature compensation point followed by negative magnetization is observed in the low zero field-cooled curves of the sample. This may be due to several reasons like formation of canted ferromagnetic domains or anti-parallel spins and clusters that are separated by anti-phase boundary, residual trapped magnetic field in solenoids of the superconducting magnetometer, presence of diamagnetic $Co^{3+}$ ions, as well as dominating anti-ferromagnetic clusters of transition metal ion pairs and anti-ferromagnetically coupled rare earth magnetic moments at low temperature. Further, from M-T curve it was concluded that ECMO is completely paramagnetic at the higher temperature (above 200 K). There is a sudden increase in magnetization below the temperature 124.5 K (which is temperature $T_C$, where the derivative of MZFC–T curve exhibits a maximum shown in fig. 2(b)) for 1000 Oe applied field. This is a sign of magnetic transition at temperature $T_C$, and it was found that transition temperature increases as we go towards higher applied ZFC and FC M-T curves, $T_C \propto H^{2/3}$, which is shown in figure 2(b). Between temperature ranges, 124.5 to 200 K nature of magnetic state were understood by inverse DC susceptibility ($\chi$) vs. temperature curve at different applied fields. It has a slight down-turn deviation from linear variation (C-W behaviour) below 200 K ($T_G$) and above magnetic ordering temperature ~125 K. From this it was concluded that there is nucleation of small but finite sized correlated regions and clusters having short-range magnetic ordering within paramagnetic matrix which is the signature behavior of Griffiths like phase [35,36].

The isothermal magnetizations vs. field curves of ECMO at different temperature are shown in fig. 3(a). The magnetization isotherms at 2 K displays sharp steps and at higher temperature sharpness of these steps disappear. These jumps are seen only while increasing field i.e. for 0T to 5T and are absent in decreasing field i.e. for 5T to 0T due to slow spin relaxation and re-appear in subsequent negative field direction i.e. for 0T to -5T followed by a large remanence and coercivity. In some multiferroic like $Ca_3CoMnO_6$ field-induced, sharp jumps is assume to be due to spin flop transition from *E**-type magnetic ordering (↑ ↑ ↓ ↓) to the ↑ ↑ ↑ ↓ $Mn^{4+}$ (high spin state, $S = 3/2$) at first critical field $H_{C1}$ and to ↑ ↑ ↑ ↑ with $Co^{2+}$ (low spin state, $S = 1/2$) at second critical field $H_{C2}$ fields. Similarly, people were trying to explain the origin of metamagnetic transition due to the field-induced spin reorientation of Co and Mn ions in $Y_2CoMnO_6$. But the amount of discontinuity in $Y_2CoMnO_6$ was not consistent with the spin flop of $Co^{2+}$ (high spin $S = 3/2$) to the ↑ ↑ ↑ ↑ magnetic structure at second jump[37]. A similar problem can be seen in ECMO too. Other than this, there is an increase and decrease of the number and step size when we change the sweep rate of the external magnetic field. So here we can't say field-induced spin flop may be a possible origin of MMT. However, these discontinuous jumps in magnetization might be due to spin flopping and magnetic-field-induced lattice distortion (martensitic-like transition) in the sample, give rise to transition between the AFM and FM states i.e. meta-magnetic transition, obtained for certain critical value of the external magnetic field. The reason behind this martensitic like-transition is strong lattice distortion by an external magnetic field via magneto-elastic coupling giving rise to crystallographically distorted FM phase. Due to this distortion, an elastic strain develops at the FM/AFM anti-phase interface. The FM phase advances when an external magnetic field is applied, but magneto-striction energy (the energy which opposes the interfacial/martensitic strain) go up against this to stop the expansion of the FM phase.

However, while cooling the sample in zero field modes, at a lower temperature, it may be possible that the AFM ordering will lead along with little FM clusters separated by an antiferromagnetic, antiphase boundary(APB). The main reason behind the pinning of Co-O-Mn ordered FM domains in antiparallel is strong AFM Co-O-Co or Mn-O-Mn interactions across the APB would orient neighboring FM domains in antiparallel[37]. For the case, when we increase the magnetic field during the isothermal magnetization process, the FM domain will try to grow, however, due to opposing forces, there is a small change in the magnetization. At a critical field $H_{C1}$, the energy of the external field is so high so that it can overcome the magneto-striction energy related to the FM/AFM anti-phase boundary and some of the spins from AFM domain flop along the field direction, giving rise to a sudden increase in magnetization. This process leads to an increase in elastic energy and decrease in the magneto-static energy, a balance which can show the way to the system to go in another metastable state. As we further increase the external magnetic field it will reach to another critical value $H_{C2}$, so that it again flop few spins and cause the second step in isothermal magnetization. Thus overall transitions proceed by successive jumps between one metastable states to another.

For more physics underlying this, we saw the magnetization jumps by changing the sweep rate of the magnetic field. Fig. 3(c,d) is showing the first cycle of the M-H curve recorded for two different sweep rates of the applied field (60 and 15 Oe/sec) with ZFC condition. For 60 Oe/s sweep rate the jump is observed at a critical field ($H_C$) ~ 28660 Oe and when we decrease the sweep rate to 15 Oe/s, it gets shifted towards the higher field $H_C$ ~ 29950. High sweep rates act as a triggering signal and strain-induced propagates quickly and converts to the FM phase. While for low sweep rate, the lattice has sufficient time to settle in the induced strains. As we increase the magnitude of the sweeping field more and more volume fraction of the sample converts to FM phase at the expense of AFM background. From figure 3(b) it is clear that at low sweep rate and high magnitude of the sweeping field we get more but small steps. It might be because of at higher field (It can be seen by taking FC isothermal magnetization data, here we have shown in fig. 3(c) a isothermal magnetization curve by cooling the sample in 3T DC magnetic field.) we have large amount of FM phase (while at low field there is dominant AFM phase) and large induced strain at AFM/FM anti-phase boundary. At low sweep rate sample has sufficient time to get a feel for increasing field and give rise to a large number of small steps while at high sweep rate system it has lesser time to acclimatize the field-effect and hence gives a small number of large steps. Another interesting feature was seen by taking the isothermal magnetization curve by cooling the sample in 1T DC magnetic field, just after taking isothermal magnetization curve by cooling the sample in 3T DC magnetic field. Here we are able to see that there is large magnetization even at zero field, which is induced due to sample was cooled in large DC field earlier than 1T, which means that it has a large memory effect.

For more information regarding spin-structure, we have gone for ac susceptibility. Fig. 4(a,b) is showing the temperature dependence of in-phase ($\chi'$) and out of phase component ($\chi''$) of AC susceptibility at different frequencies and constant AC field of 3 Oe without any DC biased. We observed unusual sharp peaks in ac susceptibility near $T_C$ around 124 K which might be due to magnetic ordering or spin glass transition. Almost peaks do not shift with the frequency which is disapproval of what is expected i.e. a typical spin glasses behavior is ruled out. Since peaks are sharp and narrow, so we can say all spins of all ions of ECMO relax with almost equal relaxation time. Further, from the investigation of out of phase component of ac susceptibility, we conclude that there are small peaks showing short-range ordering earlier than long-range ordering. More interestingly, prior to short-range ordering, there is a slight deviation in ac susceptibility from its zero value which is in favor of signature behavior of a typical Griffiths like phase.

Fig. 4(c,d) is showing the temperature dependence of in-phase ($\chi'$) and out of phase component ($\chi''$) of ac susceptibility with different DC biased at a constant frequency of 500Hz and constant AC field of 3 Oe. Here, again similar unusual sharp peak in AC susceptibility was observed near $T_C$, more interestingly this peak suppressed, broadens and splits into two maxima with increasing under influence of DC bias. The peak which is at the higher temperature, $T_{high}$ shifts towards the higher temperature side on increasing DC bias. It is to be noted that there is a similar variation in $T_C$ obtained by the M–T curve which shifts towards the high-temperature side as we go for high field ZFC/FC curves. This is consistent with the shift of the $T_{high}$ with the external dc magnetic field, which is recognized to be the presence of critical fluctuations associated with a continuous transition to a ferromagnetic state. While the peak which is at a lower temperature, $T_{low}$ shifts towards the lower temperature side on increasing DC bias. In phase component of AC susceptibility is expected to be constant below the PM to FM transition temperature in FM material. However, in a few magnetic oxide systems in which the finite size of FM and AFM clusters coexist, $\chi'(T)$ decreases below the transition. This phenomenon is known as Hopkinson effect, something which is related to magnetic anisotropy/inhomogeneity. Change in magnetic anisotropy arises on decreasing temperature due to continuous change in size and shape of FM clusters below the transition. Hopkinson effect is attributed to the competing spin-spin correlations and large

anisotropy field compared to the measurement field which can be explained by the following relation,

$$\chi'(T) \propto M_S^2(T)/K(T) \qquad \text{---(3)}$$

where $M_S(T)$ is the spontaneous magnetization and K(T) is the anisotropy energy of the system at a particular temperature T. At a temperature slightly below $T_C$ the FM spin-spin correlations become stronger and the go-ahead to a large spontaneous magnetization ($M_S$). Further, $M_S(T)$ is almost constant below the transition temperature and hence the decrease in $\chi'(T)$ might be associated with a simultaneous increase in K(T). However, it may also be possible that the observed sharp peak could be the result of both presences of ASD's and local magnetic frustration due to competing interactions between Co-O-Mn for FM and AFM in ECMO.

More interestingly, by comparing out of phase component of AC susceptibility without DC biased and with DC biased at a constant frequency and AC field, we can see an important difference. Firstly, the smaller peaks showing short-range ordering does not appear under the application of even small dc field which is again provable of Griffiths like phase. Second, in favor of a typical spin-glass, the peak gets suppressed and peak temperature ($T_P$) shifted towards lower temperature with an application of the DC magnetic field as the applied field gets in the way to the spins to freeze. Again we have fitted $T_P$(K) for different DC fields, $H^{2/3}$. It follows the de Almeida-Thouless (AT) line,

$$T_P(H) = T_P(0)(1-AH^{2/3}) \qquad \text{---(4)}$$

Where A is some constant, $T_P(0)$ and $T_P(H)$ is the value of $T_P$ in the absence of DC bias and with DC bias, and thus it is in favor of the existence of the volume spin-glass-like behavior[38-40]. And this is also agreed with what is expected i.e. mean-field scenarios which are already reported[35]. Further, fig. 4(e,f) is showing the thermal variation of AC susceptibility, at a fixed AC field =3 Oe and DC field = 750 Oe with different AC frequency. Fig. 4(e) shows that the peak of the in-phase component of AC susceptibility which is at lower temperature side gets incite and slight shift towards lower temperature at a lower frequency. Since spins can follow even small frequency so we can say these peaks might be due to slow spin relaxation. In contrast, the origin of the higher temperature side peak is something different, which might be due to what we expected i.e. long-range ordering and of course might be due to magnetic transition.

## XPS STUDY

Further, to explore magnetic properties much more, we need to look at electronic structures and valence states of elements present in the system. For this, we have carried out X-ray photoemission spectroscopy (XPS) measurement, which is shown systematically in Fig. 5(a-h). The wide-surface-scan XPS spectra of the ECMO sample is shown in Fig. 5(a) and all peaks are allocated according to the National Institute of Standards and Technology (NIST) database. From here we can see the existence of Eu, Co, Mn and O at the surface. Moreover, the only presence of the C impurity at 284.7eV was detected in the spectrum, as it always gets absorbed from the air at the surface of the sample. The core-level Eu3d XPS spectrum region of ECMO is shown in Fig. 5(b) and peaks were fitted using XPSPEAK41 software. The binding energies of Eu3d core-level have two intense peaks $Eu3d_{3/2}$ (1163.8 eV) and $Eu3d_{5/2}$ (1133.9 eV) separated by 29.9 eV, as a result of strong spin-orbit coupling. Fitting of Eu3d peaks confirms the existence of $Eu^{3+}$ state (~94.3%) due to the $3d4f^6$ final state configuration. Moreover, the $Eu3d_{3/2}$ core level shows a satellite peak around 1156 eV which corresponds to the $Eu^{2+}$ state (~5.7%) which is attributable to the $3d4f^7$ final state configuration[41]. Additionally, a low intense peak was observed above the Eud3/2 peak manifest as "Multiplet_$Eu^{3+}$" around 1142.5 eV, which is apparently related to the multiplet effect due to the coupling between the open-shell and a core electron vacancy and is reported as a satellite peak[42]. Calculation of area under the corresponding peaks fitted curve give the relative concentration of $Eu^{3+}$ states about 94.3% and that of $Eu^{2+}$ states about 5.7% respectively on the surface. Reason behind the presence of $Eu^{2+}$ states might be due to oxygen vacancies or due to formation of of some $OH^-$ ions which is shown as $O^{2-}$ loss in O1s XPS spectra(Fig 5(g)).

The core-level Co2p peak fitted XPS spectrum of ECMO is shown in Fig. 5(c). Due to spin-orbit coupling, it has two narrow intense peaks centered around 779.7 eV ($2p_{3/2}$) and around 795.4 eV ($2p_{1/2}$) corresponds to $Co^{3+}$ states[43, 44]. And other due to deconvolution, relatively broad intense peaks of $Co^{2+}$ states centered around 781.1 eV corresponds to $2p_{3/2}$ and around 797.1 eV corresponds to $2p_{1/2}$. Additionally, two peaks around 786.2 eV & 802.5 eV were observed which corresponds to shake-up satellite peak and they are very sensitive to valence states of cation, the coordination number of ligands, etc. The relative concentration of $Co^{2+}$ states was calculated to be about 59% and that of $Co^{3+}$ states about 41% respectively.

The core-level Co3s peak fitted XPS spectrum is showing a intense peak around 100.6 eV (Fig. 5(e)) and a buried peak about 106.2 eV which arises due to the exchange interactions among the core hole and the open 3d shell[45, 46].

Similarly, the core-level Mn2p spin-orbit coupling spectra show two peaks corresponding to $Mn2p_{3/2}$ and $Mn2p_{1/2}$ respectively with a doublet separation of ~12 eV (Fig. 5(d)). XPS peak fitting of the Mn2p region confirm two mix valence states of Mn, i.e. $Mn^{3+}$ (~34%) and $Mn^{4+}$(~66%) as $Mn2p_{3/2}$ are deconvoluted into two peaks centered around 641.3 eV ($Mn^{3+}$) and 642.4 eV ($Mn^{4+}$) while the $Mn2p_{1/2}$ deconvoluted peaks are around 653.3 eV ($Mn^{3+}$) and 654.4 eV ($Mn^{4+}$)[47]. Further, these Mn valence states were verified more quantitatively by taking Mn3s XPS spectra. Fig. 5(f) is showing peak fitted XPS spectrum of Mn3s core-level. Here, we observed two peaks centered around 83.76 eV and 88.46 eV because of exchange splitting with and amount of exchange splitting energy ($\Delta E_{3s}$) for our sample is about 4.7 eV. $\Delta E_{3s}$ is directly in a linear relationship with Mn valence state ($V_{Mn}$) by the relation[48],

$$\mathbf{V_{Mn} = 9.67 - 1.27\ \Delta E_{3s}\ (in\ eV)} \qquad \text{--- (5)}$$

It gives Mn valence states about 3.7+ and hence supports mix valency which is seen in Mn2p XPS spectra.

## CONCLUSION

In summary, it can be concluded that DP ECMO, prepared by the conventional solid-state reaction method has a strong magnetic frustration due to different competing Co-O-Mn exchange interactions. From inverse DC susceptibility it can be seen that there is the signature behavior of Griffiths like phase. From the isothermal magnetization curve, it was concluded that ECMO shows an MMT at a lower temperature due to the pinning of FM domains by antiferromagnetic antiphase boundaries. AC susceptibility measurement shows a sharp peak in the AC susceptibility near $T_C$. In phase component of AC susceptibility gets broaden and split into two peaks under DC biased which is similar to: (1) Hopkinson peak and might be due to the competing spin-spin correlations and large anisotropy field. (2)The peak which is at lower temperature side take over at lower frequency so it may be due unusual slow spin relaxation. Again from out of phase component of AC susceptibility with DC biased showed a shifting of peak temperature, which proves the existence of the volume spin-glass-like behavior and fallow mean-field scenarios. Analysis of XPS measurement shows the existence of mixed valence states of transition metal ions (Co and Mn).

## ACKNOWLEDGEMENTS

The authors thank to the Central Instrumentation Facility Centre, Indian Institute of Technology (BHU) for providing the low temperature magnetic measurements facility.

## Figures Caption:

**Fig. 1**: X-Ray diffraction pattern with Lebail refinement for ECMO.

**Fig. 2**: (a) The thermal variation of magnetization curves at different applied ZFC and FC condition with a inset showing inverse DC susceptibility as a function of temperature. (b) Showing derivative of MZFC–T curve and inset is showing variation of $T_C$ & $T_a$ with H.

**Fig. 3:** The isothermal magnetizations vs. field curves of ECMO at (a) different temperature, (b) Same temperature without and with 3T, 1T field cooled, (c,d) Same temperature but with different sweep rate of 60 & 15 Oe/s taken in first quadrant only.

**Fig. 4:** The thermal variation of ac susceptibility $\chi'$ and $\chi''$ at (a&b) different frequency without DC, (c&d) constant frequency and with different DC bais along with inset showing AT line, (e&f) different frequency and with constant DC bais, H= 750 Oe.

**Fig. 5:** (a-h) are showing wide-range-surface scan & core-level X-ray photoemission spectroscopy (XPS) of different elements present on the surface of ECMO and (i) is showing valence band spectrum.

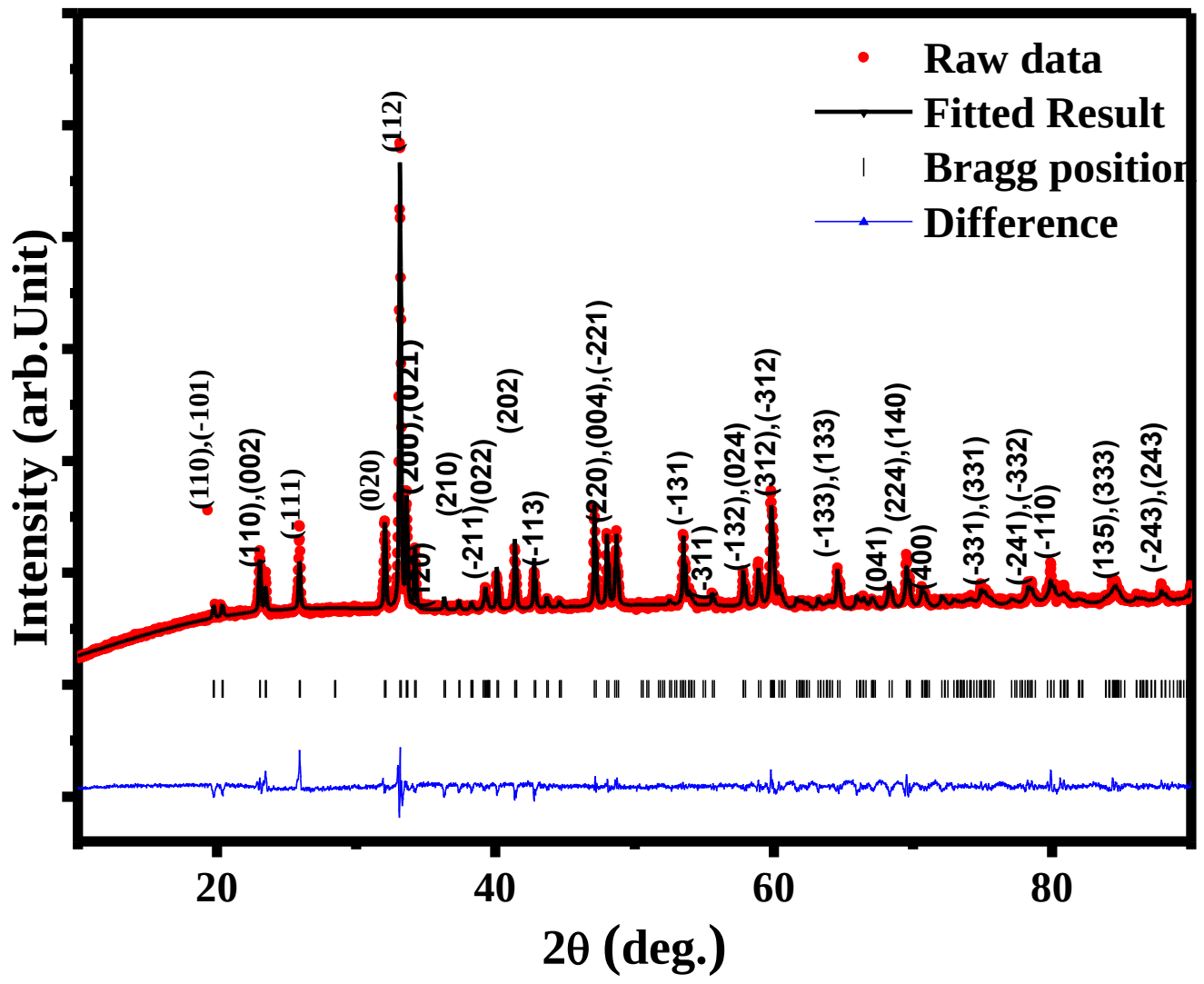

Figure 1

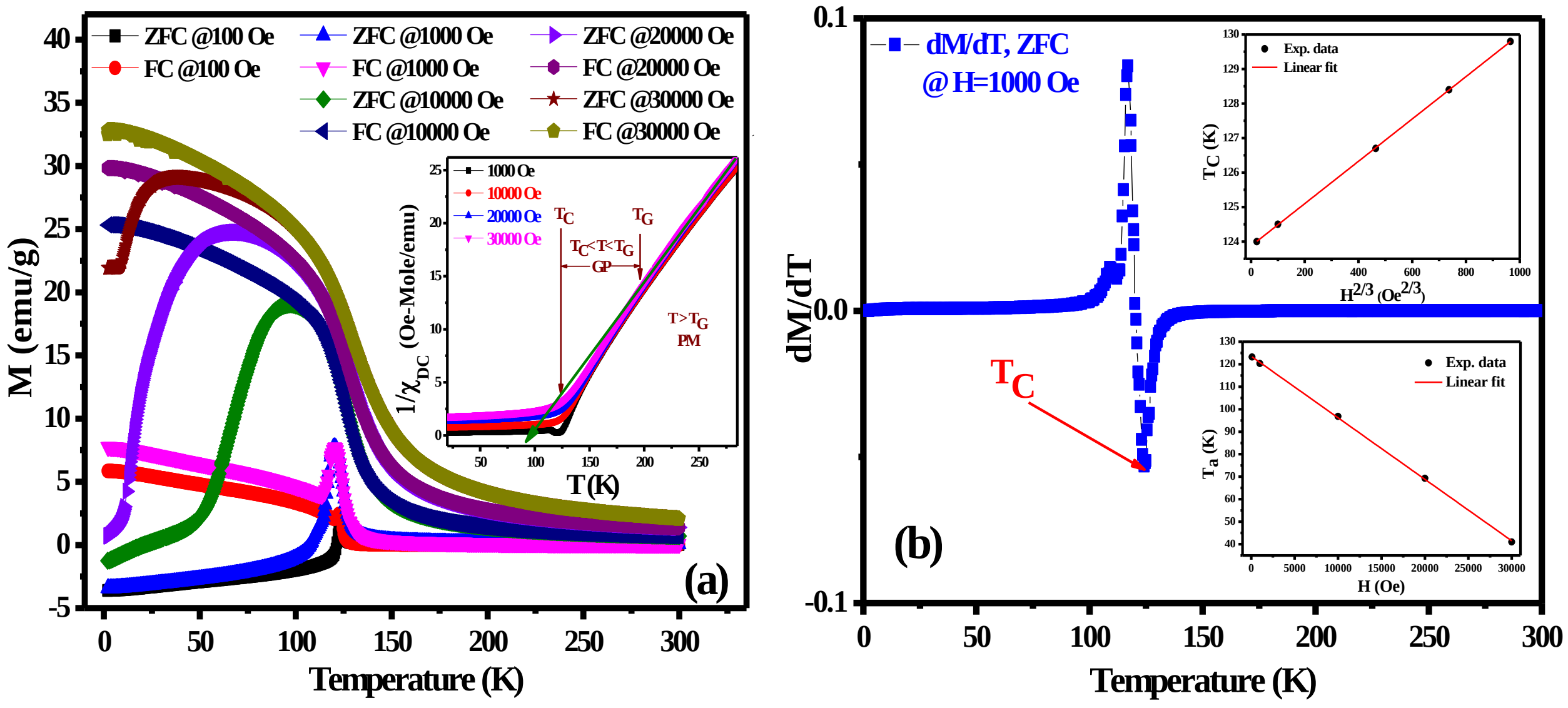

Figure 2

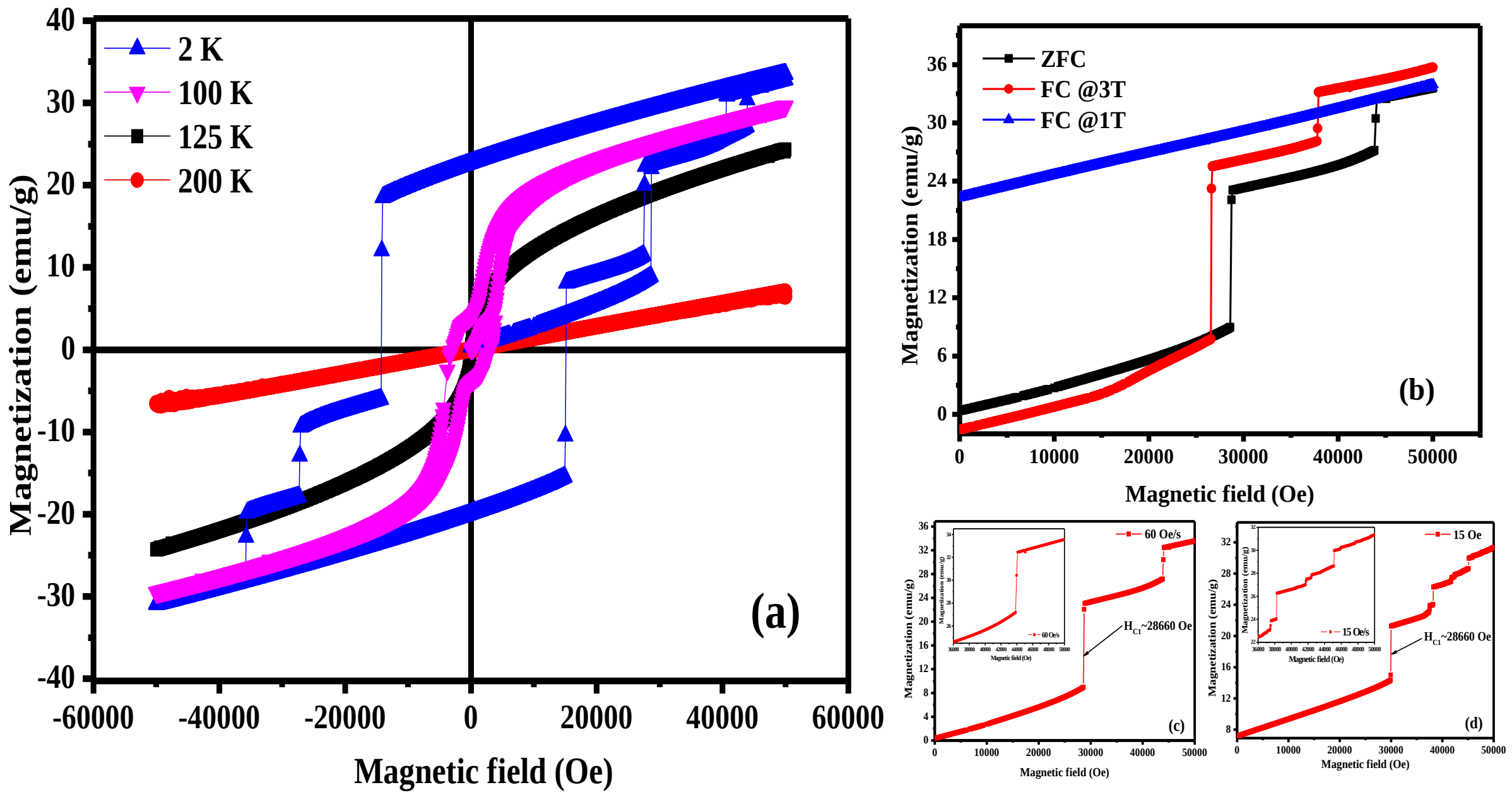

## Figure 3

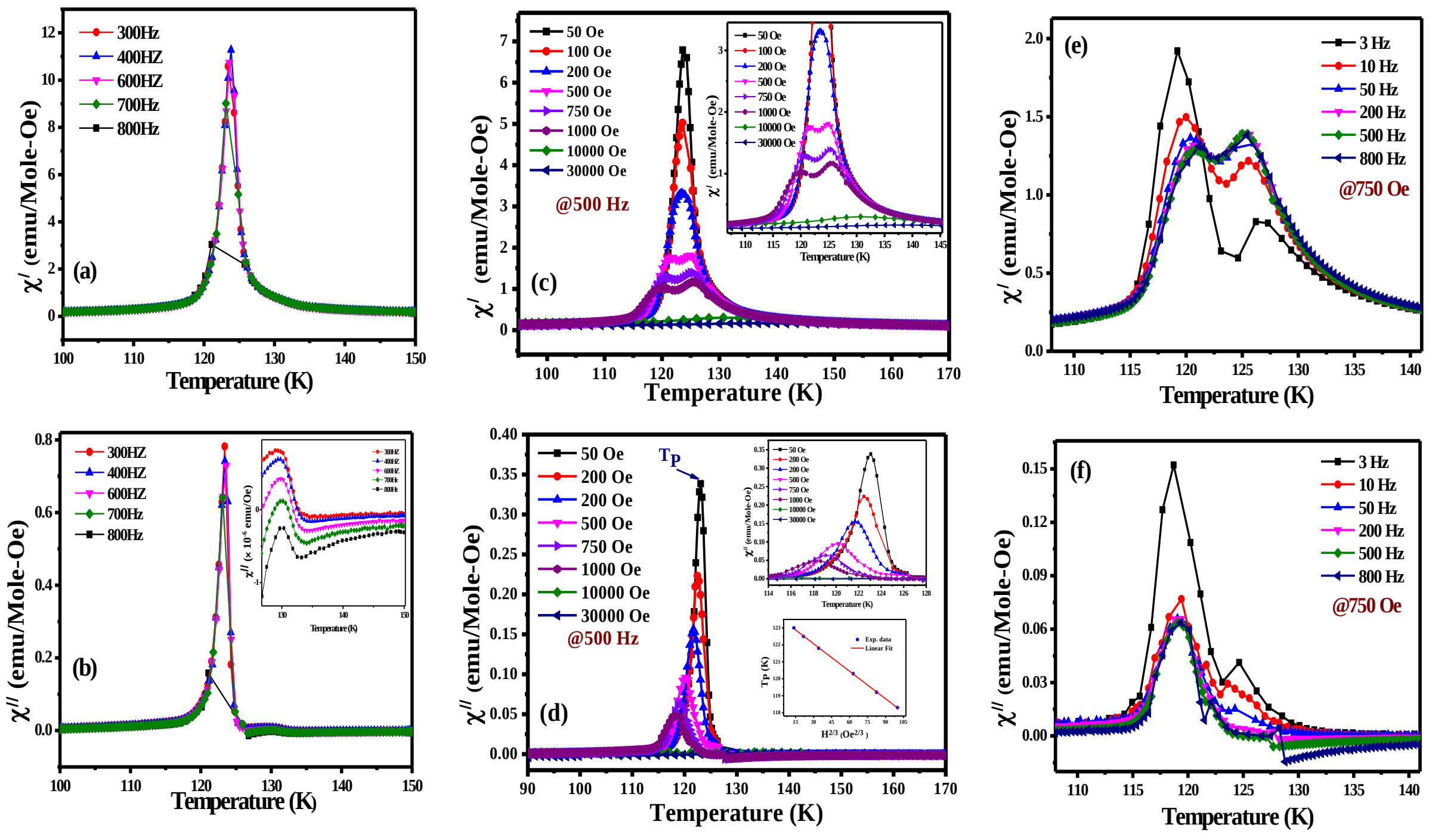

## Figure 4

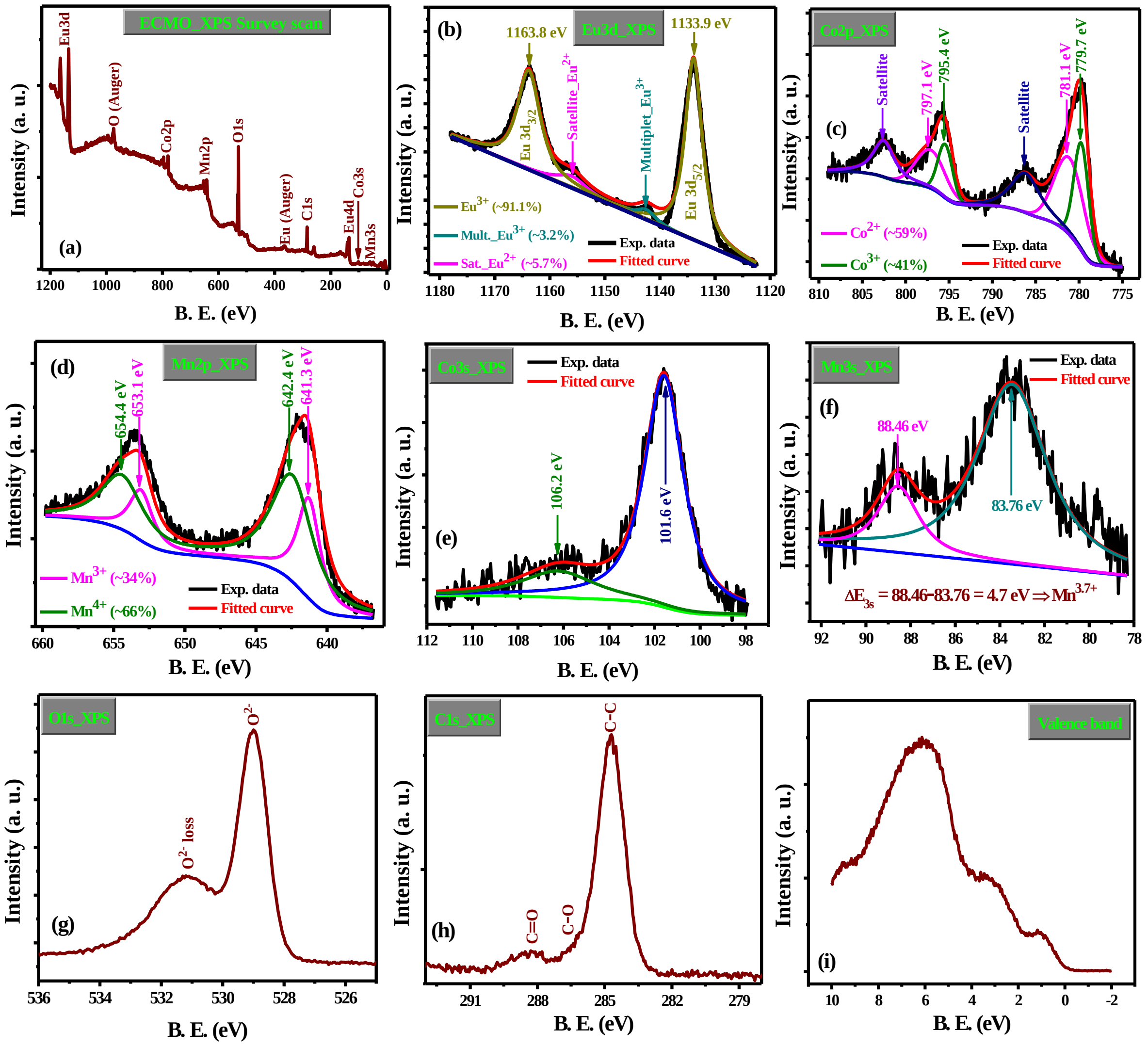

ECMO_XPS Survey scan
(a)
Eu3d
O (Auger)
Co2p
Mn2p
O1s
Eu (Auger)
C1s
Eu4d
Co3s
Mn3s
(b)
Eu3d_XPS
1163.8 eV
1133.9 eV
Eu 3d3/2
Eu 3d5/2
Satellite_Eu2+
Multiplet_Eu3+
Eu3+ (~91.1%)
Mult._Eu3+ (~3.2%)
Sat._Eu2+ (~5.7%)
Exp. data
Fitted curve
(c)
Co2p_XPS
Satellite
797.1 eV
795.4 eV
Satellite
781.1 eV
779.7 eV
Co2+ (~59%)
Co3+ (~41%)
(d)
Mn2p_XPS
654.4 eV
653.1 eV
642.4 eV
641.3 eV
Mn3+ (~34%)
Mn4+ (~66%)
(e)
Co3s_XPS
106.2 eV
101.6 eV
(f)
Mn3s_XPS
88.46 eV
83.76 eV
ΔE3s = 88.46-83.76 = 4.7 eV ⇒ Mn3.7+
(g)
O1s_XPS
O2- loss
O2-
(h)
C1s_XPS
C=O
C-O
C-C
(i)
Valence band
Intensity (a. u.)
B. E. (eV)

# Figure 5